\documentclass[aps,twocolumn,showpacs]{revtex4}
\usepackage[dvips]{graphicx} \textheight 25cm
\begin{document}

\title{Spin density in frustrated magnets under
mechanical stress: Mn-based antiperovskites}

\author{Pavel Lukashev}
\affiliation{Department of Physics, University of Nebraska at Omaha,
Nebraska 68188}

\author{Renat F. Sabirianov}

\affiliation{Department of Physics, University of Nebraska at Omaha,
Nebraska 68188}

\author{Kirill Belashchenko}

\affiliation{Department of Physics and Astronomy, University of
Nebraska-Lincoln, Nebraska 68588}

\date{\today}

\begin{abstract}

In this paper we present results of our calculations of the
non-collinear spin density distribution in the systems with
frustrated triangular magnetic structure (Mn-based antiperovskite
compounds, Mn$_{3}$AN (A=Ga, Zn)) in the ground state and under
external mechanical strain. We show that the spin density in the
(111)-plane of the unit cell forms a "domain" structure around each
atomic site but it has a more complex structure than the uniform
distribution of the rigid spin model, i.e. Mn atoms in the
(111)-plane form \textit{non-uniform} "spin clouds", with the shape
and size of these "domains" being function of strain. We show that
both magnitude and direction of the spin density change under
compressive and tensile strains, and the orientation of "spin
domains" correlates with the reversal of the strain, i.e. switching
compressive to tensile strain (and vice versa) results in "reversal"
of the domains. We present analysis for the intra-atomic
spin-exchange interaction and the way it affects the spin density
distribution. In particular, we show that the spin density inside
the atomic sphere in the system under mechanical stress depends on
the degree of localization of electronic states.

\end{abstract}
\maketitle

\section{1. Introduction}

In magnetic materials the local magnetic moments (LMM) are generally
due to the d- and f- electrons. These electrons are localized, for
example for 3d-metals the radius of d-orbitals is less than 1\AA
(\textit{localized region (LR)}). On the other hand, s- and
p-orbitals in 3d-metals are less localized or even delocalized
(\textit{delocalized region (DR)}). Thus, there are regions of high
spin density (SD) in the LR near the nucleus and low SD further away
in the DR. Despite the inhomogeneity of SD in magnetic materials, it
is frequently described within the quasispin approximation
\cite{Liu_quasispin}, according to which the direction of the SD
around each atom is taken constant within atomic sphere (AS) or
polyhedron. This approach works better for the systems with strong
spin-exchange coupling because the strong exchange between spins
ensures that the SD direction stays almost uniform. Yet, for some
systems it is essential to consider detailed distribution of
magnetization axis as this may reveal additional important features
of the magnetic configurations \cite{Nordstrom, Hafner} overlooked
within the \textit{rigid spin approximation} (RSA). In this paper we
examine the SD distribution of the systems with frustrated
magnetization, in particular Mn$_{3}$AN antiperovskites, using fully
unconstrained SD functional without any assumption on the uniformity
of SD around each atom.

Many interesting phenomena in magnetic materials are due to the
interplay of magnetic and structural degrees of freedom because the
SD is sensitive to mechanical deformations such as compression,
in-plane biaxial strain, etc. In particular, such phenomena based on
magneto-mechanical coupling have been reported for the Mn-based
antiperovskites, - for example piezomagnetic effect \cite{Lukashev},
a giant magnetoresistance of more than 10\% in pulsed magnetic
fields \cite{Kamashima}, invar effect (or even negative thermal
expansion) \cite{Takenaka}. While the SD evolves under mechanical
stress, the RSA assumes that its direction is changing uniformly
inside of the AS. Yet, it is not obvious that in the excited states
the spin moments in LR and DR rotate to the same angle. \textit{We
demonstrate in this paper that the spin directions in these two
region may rotate in opposite directions}.

\begin{figure}[h]
\begin{center}\vskip 0.5 cm
\includegraphics[width=6cm]{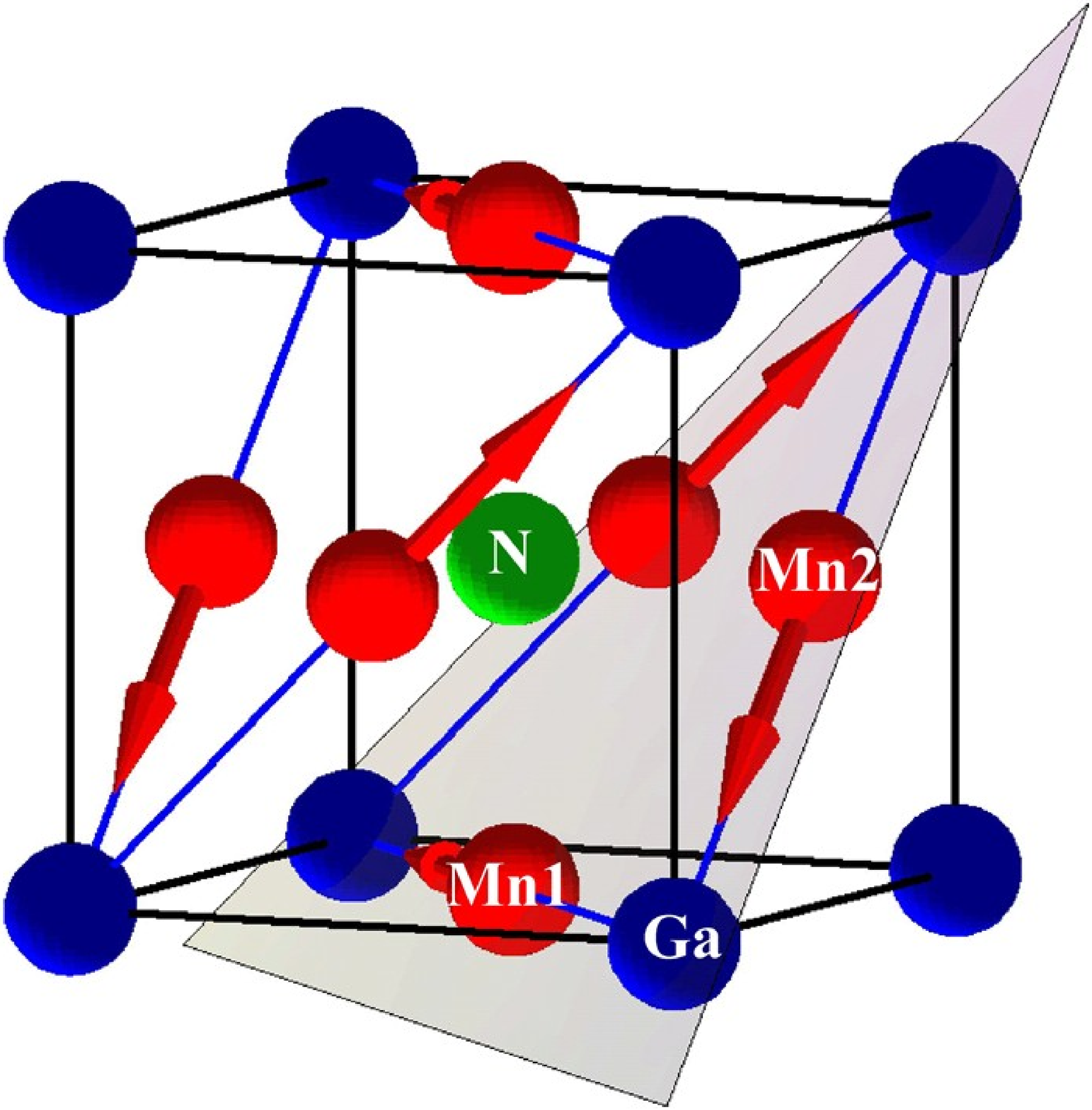}
\includegraphics[width=7.25cm]{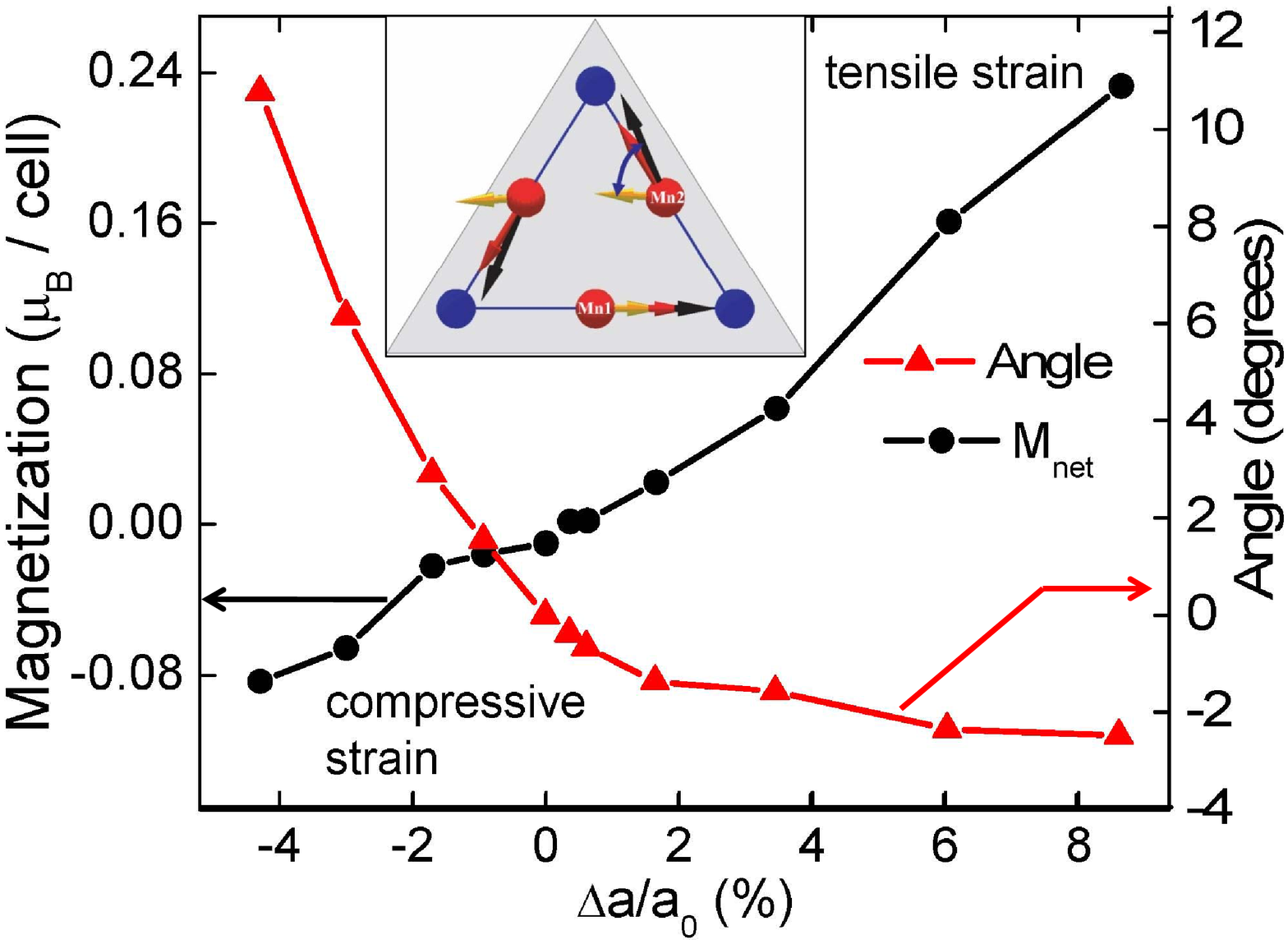}
\caption{(color online) Top panel: non-collinear $\Gamma$$^{5g}$
structure, unit cell. Blue spheres - Ga atoms; red spheres - Mn
atoms; green sphere - N atom. Arrows represent LMMs of Mn atoms.
Bottom panel: the net magnetic moment per unit cell and the rotation
angle of Mn2 LMMs as functions of biaxial strain ($\Delta$a/a (\%))
in Mn$_{3}$ZnN. Inset on bottom panel: schematic view of the
variation of Mn LMMs for Mn$_{3}$AN in (111)-plane as a function of
strain. Yellow arrows - compressive strain, black arrows - tensile
strain, red arrows - ground state.
\label{ground-state}}
\end{center}
\end{figure}

The ground state of Mn$_{3}$AN is shown in Fig.~\ref{ground-state}
(top panel). It forms an antiperovskite crystal structure with cubic
space group \textit{Pm3m}. The orientation of the LMMs of the Mn
atoms forms a non-collinear, $\Gamma$$^{5g}$ structure in the
Bertaut's classification \cite{Bertaut}. This is a structure with
the spins on the (111)-plane with clockwise or counterclockwise spin
configurations, such that the spin moments in the plane are
canceling each other. This somewhat puzzling magnetic structure
raises a natural question on how one treats the SD distribution in
the system. In our previous work \cite{Lukashev} we analyzed the
piezomagnetic properties of Mn$_{3}$AN without presenting any
details on the SD distribution, but rather we have discussed LMMs
integrated over the atomic spheres.

The paper is organized as follows: in Section 2, we present results
of our calculations of the magnetic properties of antiperovskite
such as Mn$_{3}$ZnN under external stress; in Section 3, we present
our results on the SD distribution in the (111)-plane of the unit
cell, and we show that it has more complex structure than the one
postulated in the RSA; in Section 4 we analyze the structure of the
SD distribution in terms of the intra-atomic spin-exchange
interaction; and finally, the last section is devoted to the
conclusions.

We employ the projector augmented-wave (PAW) method originally
proposed by P.  Bl\"ochl \cite{Bloechl}. We use its implementation
by G. Kresse and D. Joubert in the Vienna \textit{ab} initio
simulation package (VASP) code \cite{Kresse} within a
Perdew-Burke-Ernzerhof (PBE) generalized gradient approximation
\cite{PBE} of the density functional theory (DFT). We use a
$12\times12\times12$ k-point sampling and the Bl\"ochl's tetrahedron
integration method \cite{Bloechl1}. We set the plane-wave cut-off
energy to 300 eV and we choose the convergence criteria for energy
of 10$^{-5}$ eV. Within this framework we perform the non-collinear
magnetic structure calculations, and we allow the LMMs to relax to
the equilibrium configuration.

\section{2. Magnetic properties of Mn-based antiperovskites under stress} 

The equilibrium parameters for the Mn$_{3}$ZnN are as follows:
3.87\AA \hspace{0.001 mm} for the lattice constant and 2.5 $\mu_{B}$
for the \textit{LMM integrated over the volume of the space-filling
AS}. The magnetic moment is extremely sensitive to the volume
change. It steadily increases from 0.5 $\mu_{B}$ to 3.7 $\mu_{B}$ as
the volume increases from the lattice parameter 3.5\AA \hspace{0.001
MM} to 4.3\AA \hspace{0.001 MM}. Application of the external strain
results in induced magnetization. As one can see from the inset on
the bottom panel of the Fig.~\ref{ground-state} the magnetic moment
of the Mn1 atom in the basal plane (formed by atoms of Mn and Ga)
preserves its direction but its magnitude decreases/increses under
compressive/tensile strain. LMMs of the Mn2 atoms in the Mn$_{2}$N
plane change both their direction and magnitude, - they rotate
towards the [110] direction when the compressive strain is applied
in such a way that they become almost opposite to the LMM of Mn1
atom at large strains. The LMMs of Mn2 atoms become more aligned
with each other (see Fig.~\ref{ground-state}, - inset on the bottom
panel). When tensile strain is applied the LMMs of Mn2 atoms rotate
in direction opposite to the one at the compressive strain and the
Mn2 moments become more anti-aligned. The magnitude of the LMMs
generally decreases with the compressive strain and increases with
the tensile strain. Thus, Mn$_3$AN (A = Ga, Zn) antiperovskite
acquires a net magnetization under biaxial strain as a result of the
rotation of LMMs and change in their magnitude. The net
magnetization is directed along the C$_{2z}$ axis (this corresponds
to a diagonal axis C$_{2a}$ in the cubic cell).
Fig.~\ref{ground-state} (bottom panel) shows magnetization per cell
(black line with solid circular data points) as a function of
biaxial strain. The strain induced magnetization is about 0.07
$\mu_B$ per unit cell for the Mn$_3$ZnN at 3\% of the compressive
strain (for Mn$_3$GaN this value is substantially larger, in
particular it is about 0.16 $\mu_B$ at 2\% of the compressive
strain). The piezomagnetic effect in Mn$_3$AN is {\em linear} and
exhibits {\em magnetization reversal} with the applied strain, i.e.
the direction of the magnetization along the C$_{2z}$ axis reverses
with the reversal of the applied strain. The rotation angle of the
LMM of Mn2 atom under biaxial strain with respect to the Mn moment
direction in the ground state as a function of strain is shown in
Fig.~\ref{ground-state} (bottom panel). The dependence is linear in
the range of $\sim\pm2\%$ of strain, and the LMMs rotate in opposite
direction if the sign of the strain is reversed.

\section{3. Spin Density Distribution}

\begin{figure}[h]
\begin{center}\vskip 0.5 cm
\includegraphics[width=7cm]{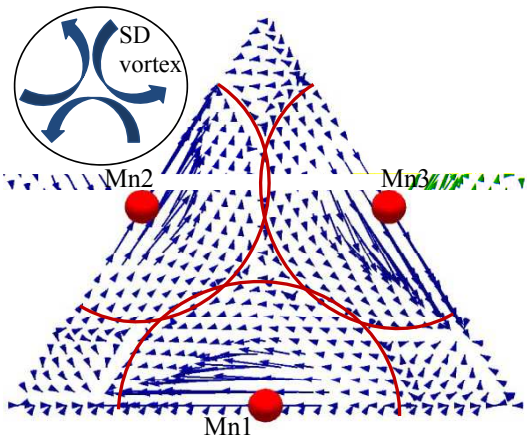}
\caption{(color online) Spin density distribution in (111)-plane for
the non-collinear $\Gamma$$^{5g}$ structure of Mn$_3$AN, unit cell.
Red spheres represent Mn atoms. Red arcs schematically highlight SD
distribution "domains" around each Mn atom. Blue arrows represent SD
direction. Inset schematically shows the orientation of the spin
density in the center of the vortex. \label{spin-dens}}
\end{center}
\end{figure}

Fig.~\ref{spin-dens} shows the SD distribution in the (111)-plane of
the unit cell for the non-collinear ground state $\Gamma$$^{5g}$
structure of Mn$_3$AN. The overall structure of the SD in the
(111)-plane resembles vortex with the center in the middle of the
triangle formed by the Mn atoms. The orientation of the spin density
in the center of the vortex is shown schematically on the inset of
the Fig.~\ref{spin-dens}. It reflects the anti-ferromagnetic nature
of the inter-atomic exchange coupling of the nearest neighbor Mn
atoms in the (111)-plane. Red arcs on the plot schematically show
the AS within which the SD direction is assumed to be uniform by the
quasispin approximation. As one can see from the plot, \textit{this
uniformity is only very approximate} as the magnetization axis
direction clearly changes within the "quasispin domain".
Particularly, in the LR (approximately inside r$<$0.7\AA
\hspace{0.02cm} sphere) spin axis direction is more or less uniform.
However, in the DR of low SD (approximately 0.7\AA
\hspace{0.02cm}$<$r$<$1.4\AA) the direction is changing
substantially (by about 30 degrees).

\begin{figure}[h]
\begin{center}\vskip 0.5 cm
\includegraphics[width=8cm]{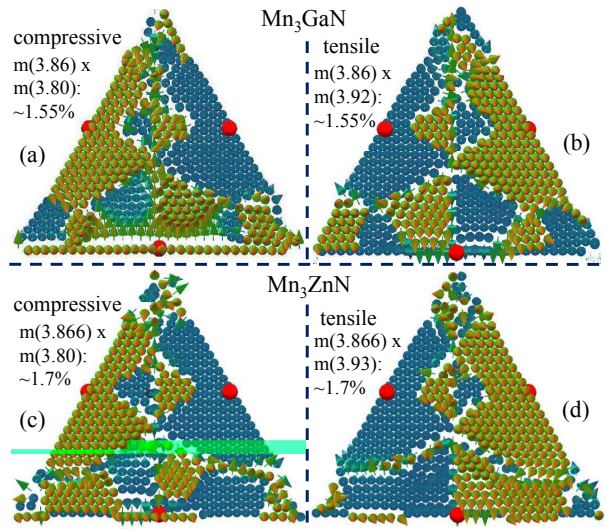}
\caption{(color online) Vector product of the local SD in the ground
state and under strain in the (111)-plane: Mn$_3$GaN (a, b) and
Mn$_3$ZnN (c, d): red spheres represent Mn atoms; green (light)
domains represent out-of-plane vector product direction; blue (dark)
domains represent into-the-plane vector product direction.
\label{Ga-Zn-magn-dens}}
\end{center}
\end{figure}

Even more dramatic effect can be observed under strain, where
regions of SD inside AS rotate in opposite directions. To better
visualize the non-uniformity of SD we plot the vector product of the
magnetizations in the ground state and under strain
(\textbf{m}(equilibrium)$\times$\textbf{m}(strain)) in the
(111)-plane (see Figure ~\ref{Ga-Zn-magn-dens}). The out of plane
arrows (green) show counterclockwise rotations, while into the plane
arrows (blue) show clockwise rotations of SD under strain. One can
clearly visualize from Figure ~\ref{Ga-Zn-magn-dens} the appearance
of domains with different direction of SD rotation. Particularly,
there are relatively large domains in DR which rotate opposite to
the domains in LR. Therefore, the SD distribution in the (111)-plane
is more complex than the ideal picture of the uniform SD around each
atom assumed by the RSA. To understand this "beyond RSA" pattern we
analyze the intra-atomic spin-exchange interaction and the way it
affects the SD around each atom.

\begin{figure}[h]
\begin{center}\vskip 0.5 cm
\includegraphics[width=7cm]{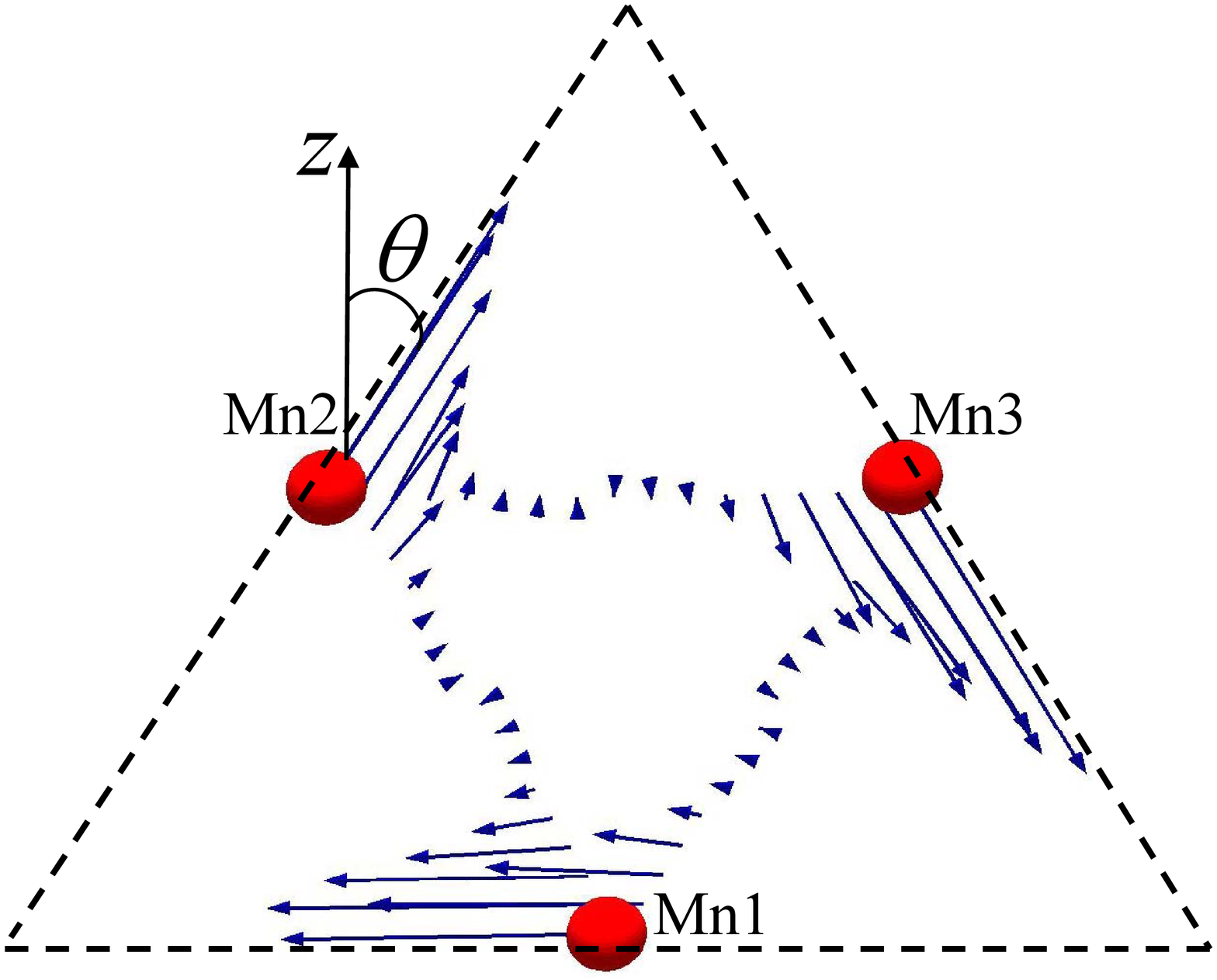}
\includegraphics[width=7cm]{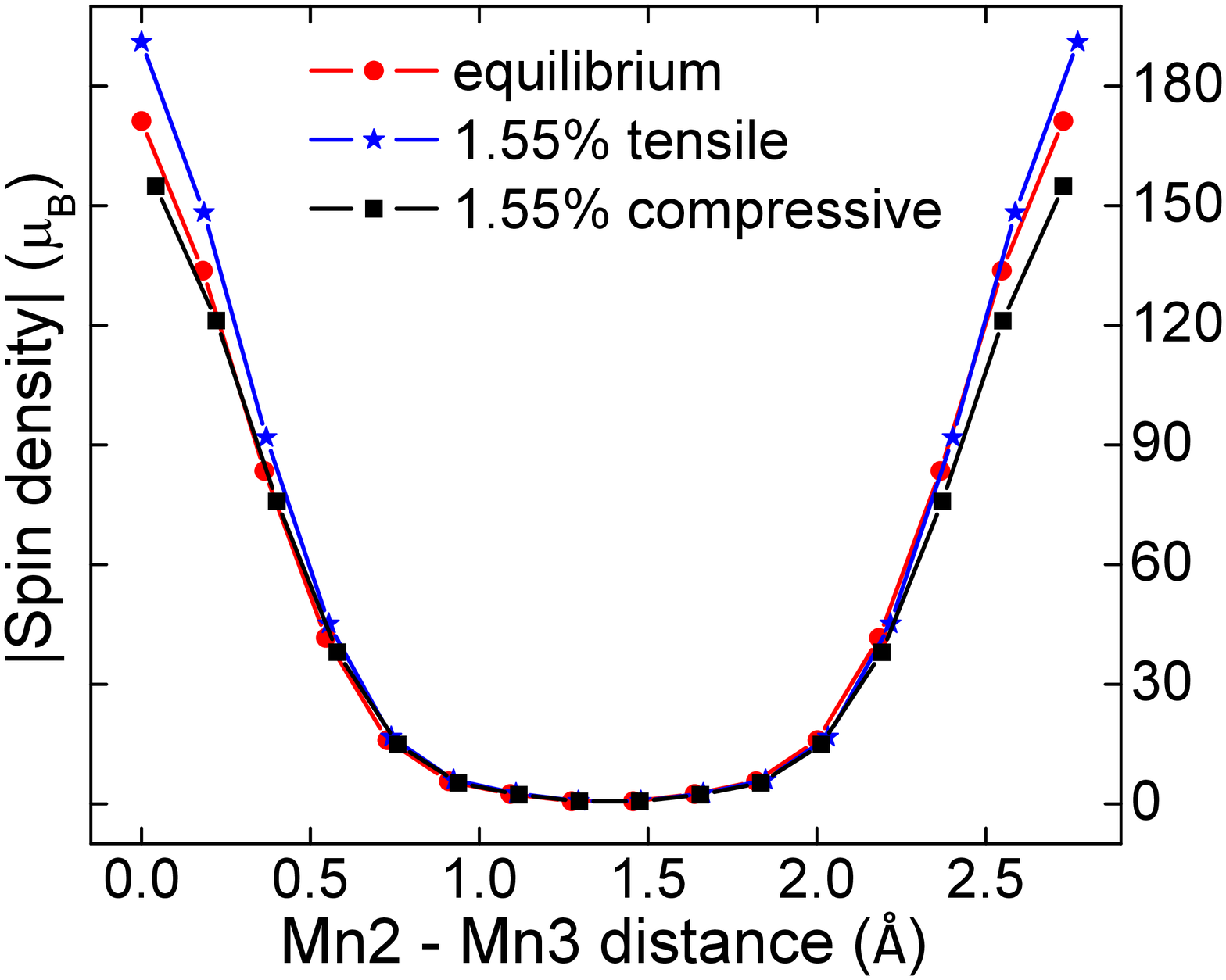}
\caption{(color online) Magnetization distribution between Mn atoms
in (111)-plane for the non-collinear $\Gamma$$^{5g}$ structure of
Mn$_3$GaN (top panel); $\mid$Spin density$\mid$ as a function of
distance between Mn2 and Mn3 (bottom panel) for ground state (red),
1.55\% tensile strain (blue), and 1.55\% compressive strain (black).
\label{MD-poloski-mn3gan}}
\end{center}
\end{figure}

\begin{figure}[h]
\begin{center}\vskip 0.5 cm
\includegraphics[width=7cm]{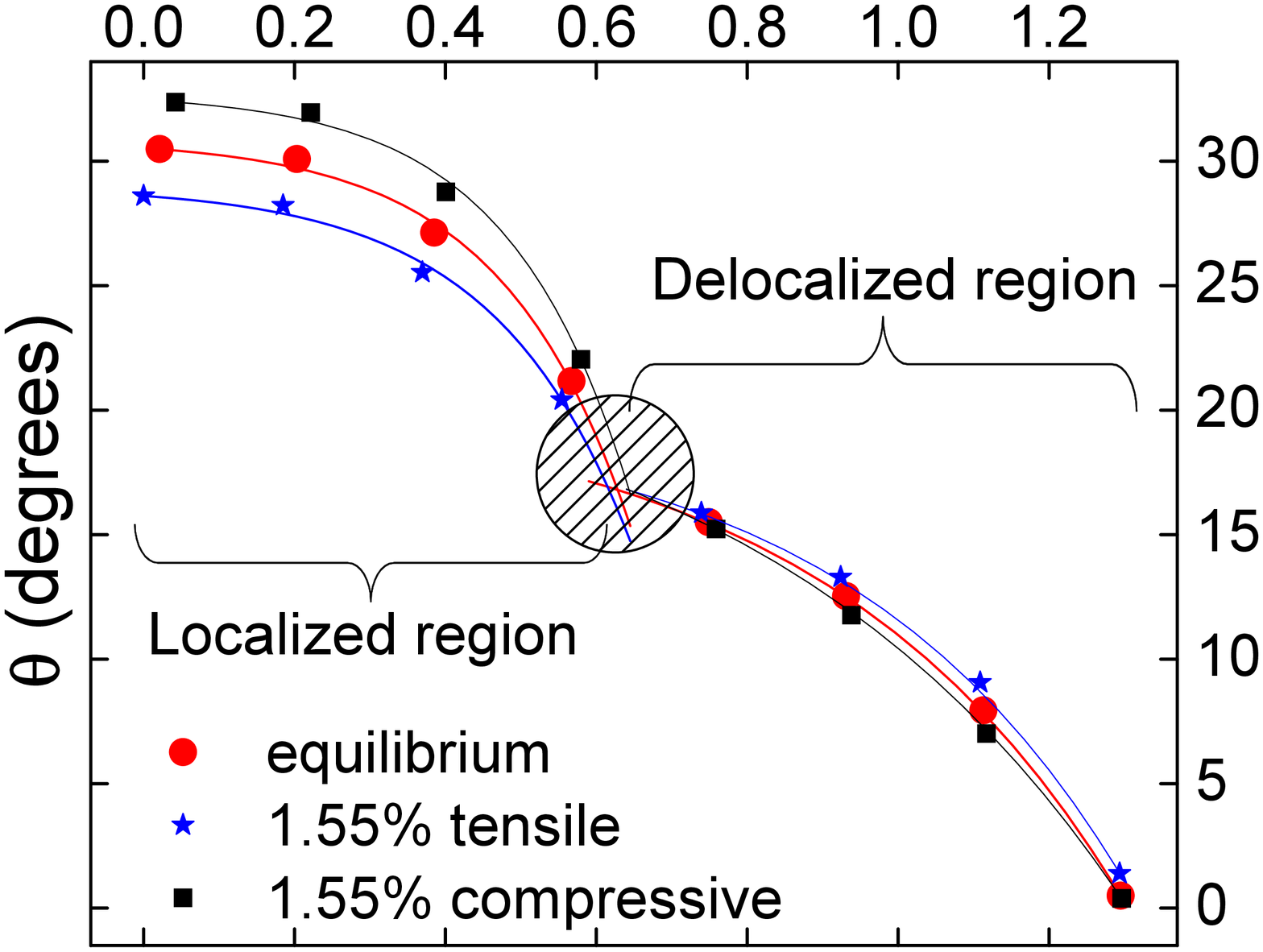}
\includegraphics[width=6.8cm]{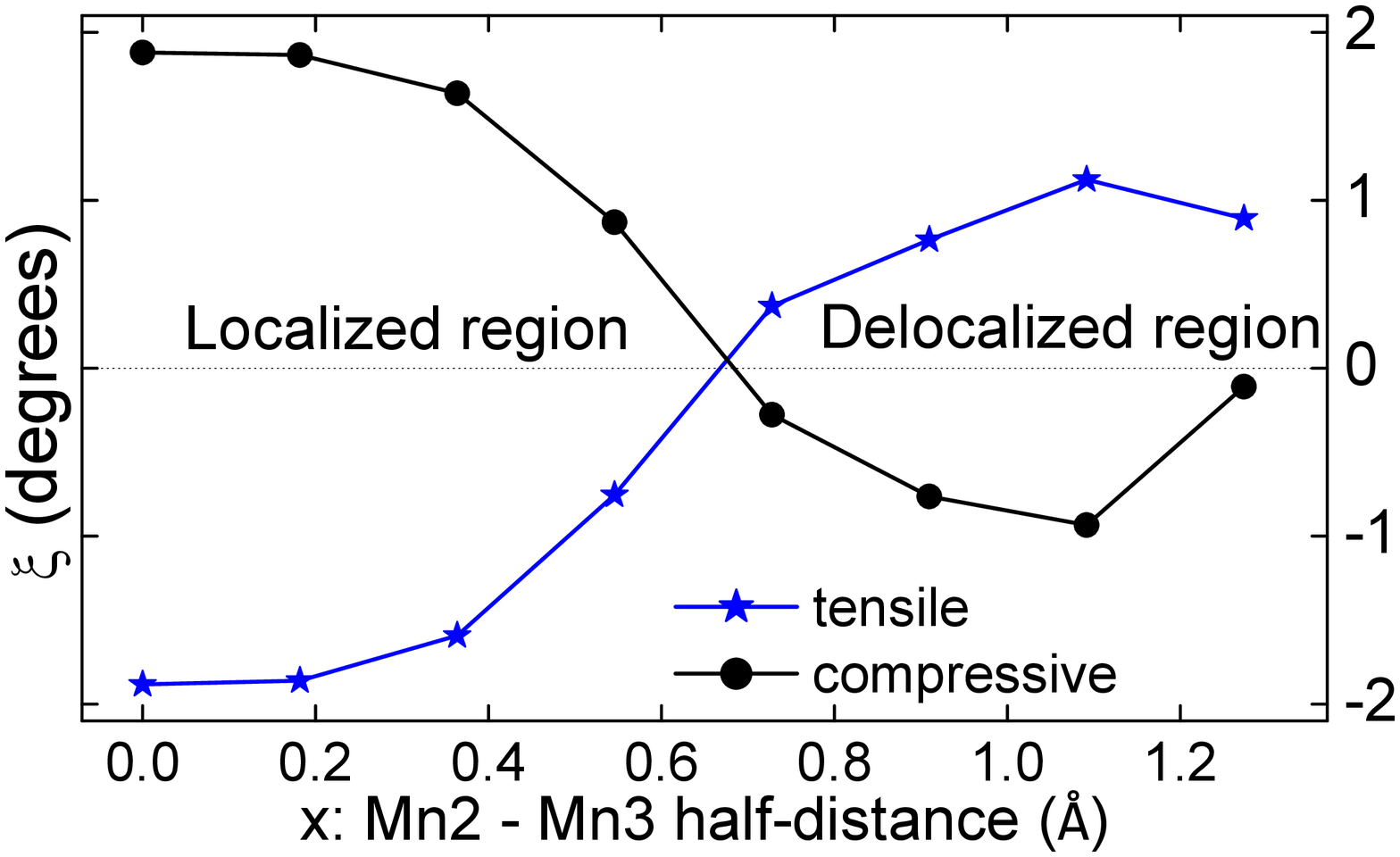}
\caption{(color online) Top panel: $\theta$ as a function of
half-distance between Mn2 and Mn3 atoms for ground state (red
spheres), 1.55\% tensile strain (blue stars), and 1.55\% compressive
strain (black squares). Bottom panel: $\xi$ as a function of
half-distance between Mn2 and Mn3 atoms. \label{exchange-angle}}
\end{center}
\end{figure}

\section{4. Discussion}

To simplify our analysis we consider SD along the line between Mn2
and Mn3 atoms, i.e. we reduce 3-dimensional problem to
1-dimensional. Fig.~\ref{MD-poloski-mn3gan} (top panel) shows the
ground state SD distribution for the Mn$_{3}$GaN along the lines
connecting Mn atoms in the (111)-plane of the unit cell. \textit{The
angle, $\theta$ between the SD direction (blue arrows on the plot)
and the z-axis (normal to the line between Mn2 and Mn3 atoms) does
not show the uniformity inside AS around each Mn atom assumed in the
RSA}. Fig.~\ref{MD-poloski-mn3gan} (bottom panel) shows the
magnitude of the SD, $\mid$\textbf{m}$\mid$ as a function of
distance between Mn2 and Mn3 atoms. Like $\theta$, the
$\mid$\textbf{m}$\mid$ also shows distinct non-uniform behavior.
Qualitatively these features of SD distribution can be interpreted
in terms of the spin-exchange interactions. For \textit{continuous}
SD distribution the spin-exchange contribution to the interaction
energy between Mn2 and Mn3 atoms can be written in terms of the spin
(Heisenberg) Hamiltonian, $H(x,x')$ \cite{Ashcroft-Mermin} as
follows:

\begin{eqnarray}
\label{exchange-energy} E_{ex.}(\theta)=-\int_{x2}^{x3}dxdx'
\underbrace{J(x,x')
S_{x}S_{x'}cos(\theta_{x}-\theta_{x'})}_{H(x,x')}\Rightarrow \nonumber \\
E_{ex.}\approx const+\frac{1}{2}\int_{x2}^{x3}dxdx'
J(x,x')s_{x}s_{x'}e^{-\frac{x+x'}{\lambda}}(\delta\theta)^{2} \nonumber \\
=const+\frac{1}{2}\int_{x2}^{x3}dx
J(x)s_{x}^{2}e^{-\frac{x}{\lambda/2}}(\delta\theta)^{2}
\end{eqnarray}
where $J(x,x')$ is the exchange parameter; $S_{x}$ and $S_{x'}$ are
spin vectors at $x$ and $x'$; $x$ and $x'$ are "spin positions"
along Mn2$\leftrightarrow$Mn3 line; $x$2 and $x$3 are coordinates of
the Mn2 and Mn3 atoms; $\theta_{x}$ and $\theta_{x'}$ are angles
between spin at $x$ and $x'$ and z-axis. Besides, here we used the
exponential behavior of the magnitude of the spin vector (see
Fig.~\ref{MD-poloski-mn3gan}, bottom panel), and
$\delta$-function-like behavior of $J(x,x')$, i.e.

\begin{eqnarray}
\label{spin-vector-exponential}
S_{x}S_{x'}=s_{x}s_{x'}e^{-\frac{x+x'}{\lambda}}
\end{eqnarray}

\begin{eqnarray}
\label{exchange-delta-like} J(x,x')s_{x'}dx'=J(x)s_{x}
\end{eqnarray}

We also assume that $x$ and $x'$ are very close to each other,
therefore $x+x'\approx 2x$.

To find the expression for the $\theta(x)$ which fits the SD
distribution shown on Fig.~\ref{MD-poloski-mn3gan} (top panel) we
have to minimize Eq.\ref{exchange-energy} w.r.t. \textit{x}, i.e. we
have to set the functional derivative of Eq.\ref{exchange-energy}
equal to zero. This results in the following differential equation
with the boundary conditions:

\begin{eqnarray}
\label{different-eq}
J(x)\cdot\ddot{\theta}(x)+\dot{\theta}(x)\cdot\dot{J}(x)=0; \nonumber \\
\theta(x2)=30^{0}; \hspace{0.2cm} \theta(x3)=-30^{0}
\end{eqnarray}
from which
\begin{eqnarray}
\label{different-eq1}
e^{-\frac{x}{\lambda/2}}\cdot\frac{d^{2}}{d^{2}x}\theta(x)+\frac{d}{dx}\theta(x)
\cdot\frac{d}{dx} e^{-\frac{x}{\lambda/2}}=0; \nonumber \\
\theta(x2)=30^{0}; \hspace{0.2cm} \theta(x3)=-30^{0}
\end{eqnarray}

We solve this equation separately for LR and DR by choosing the
value for $\lambda$ which corresponds to the best fit for the
distribution of $\theta$ as a function of distance between Mn2 and
Mn3 atoms. Results of the solutions for the ground state and
strained Mn$_{3}$GaN are summarized on Fig.~\ref{exchange-angle}
(top panel) and Table~\ref{tab:table1}.

\begin{table}
\begin{tabular}{|c|c|c|c|}
  \hline
  $J(x)$ and $\theta$ & Equilibrium & 1.55\% tens. & 1.55\% compr. \\ \hline
  $J(x)$(LR) & e$^{-x/0.17}$ & e$^{-x/0.18}$ & e$^{-x/0.16}$ \\ \hline
  $J(x)$(DR) & e$^{-x/0.40}$ & e$^{-x/0.37}$ & e$^{-x/0.50}$ \\ \hline
  $\theta$(LR) & 31-0.35$\cdot e^{5.9\cdot x}$ & 29-0.40$\cdot e^{5.6\cdot x}$ & 33-0.29$\cdot e^{6.3\cdot x}$ \\ \hline
  $\theta$(DR) & 21-0.79$\cdot e^{2.5\cdot x}$ & 20-0.57$\cdot e^{2.7\cdot x}$ & 23-1.69$\cdot e^{2\cdot x}$ \\ \hline
\end{tabular}
\caption{$J(x)$ and $\theta(x)$ for the Mn$_{3}$GaN in the ground
state and under strain for regions of localized (LR) and delocalized
(DR) electronic states.}
 \label{tab:table1}
\end{table}

Fig.~\ref{exchange-angle} (top panel) shows the change in $\theta$
for Mn$_{3}$GaN along the half-distance of the line connecting Mn2
and Mn3 atoms for the ground state, tensile strain ($\sim$1.55\%),
and compressive strain ($\sim$1.55\%) (for the other half of the
Mn2-Mn3 distance we just have to change the sign of $\theta$). As
one can see the angle $\theta$ under tensile strain for the SD
distribution in the LR is lower than the $\theta$ for the ground
state in the same region. This is consistent with the general
description of the piezomagnetic effect presented above, - see for
example the inset on the bottom panel of the
Fig.~\ref{ground-state}. Yet, in the DR the ground state $\theta$
becomes smaller, right up to the point when $\theta$ changes the
sign. Under compressive strain the $\theta$ is larger for the SD
distribution in the LR than the $\theta$ for the ground state in the
same region (again, consistent with the description of the
piezomagnetism presented above). Closer to the DR the difference in
angles for the system under compressive strain and for the system in
equilibrium decreases and finally changes its sign, after which the
$\theta$ for the ground state becomes larger than the $\theta$ for
the compressed system. But for both signs of the strain (tensile and
compressive) the magnitude of the difference in the $\theta$ for the
strained and equilibrium states
(\textit{$\xi\stackrel{\mathrm{def}}{=}\theta$(under strain) -
$\theta$(equilibrium)$\mid_{\theta\geq 0 or \theta\leq 0}$}) is
larger in the LR compared to the DR (see Fig.~\ref{exchange-angle}
(bottom panel)). This $\xi$ sign changing mechanism is obviously in
agreement with the SD distribution picture shown on the
Fig.~\ref{Ga-Zn-magn-dens} (see for example (a) and (b), - both show
the reversal of the sign of the
\textbf{m}(equil.)$\times$\textbf{m}(strain) at the border of the
LR/DR domains. \textit{These "additional spin domains" in DR cannot
be revealed within the RSA framework}. The sign of $\xi$ is changing
because the strain affects the overlap of the wave functions
(spin-exchange interaction). For example, when tensile strain is
applied the distance between Mn2 and Mn3 atoms increases which
results in increased localization of the electronic states in the LR
(in $J(x)$ $\lambda$ increases from 0.17 to 0.18, - see
Table~\ref{tab:table1}). At the same time, $J(x)$ of delocalized
states in the DR becomes weaker ($\lambda$ decreases from 0.40 to
0.37). This decrease in $J(x)$ in the DR results in reversal of the
$\xi$ sign, which in its turn leads to appearance of the additional
"domains" on Fig.~\ref{Ga-Zn-magn-dens}. On the other hand, when
compressive strain is applied, the states in the LR become less
localized ($\lambda$ decreases from 0.17 to 0.16) while the $J(x)$
in DR becomes stronger ($\lambda$ increases from 0.40 to 0.50). This
$J(x)$ increase in DR under compressive strain also results in
reversal of the $\xi$ sign (as in the case of the tensile strain the
spins in LR and DR turn in different directions). All the above
arguments based on our model present general idea of the mechanisms
responsible for the "beyond RSA" picture.

\section{6. Conclusion}

We have presented results of our calculations on the spin density
distribution of the Mn$_{3}$AN in the ground state and under
external bi-axial strain. In both cases the spin density
demonstrates non-uniform features such as appearance of additional
spin domains in the (111)-plane not revealed within the quasispin
approximation. The orientation of the spin density depends on the
direction of the strain. We have presented explanation of the
non-uniform distribution of the spin density in terms of the
intra-atomic spin-exchange interaction. In particular we have shown
that the spin density distribution depends on the degree of
localization of the electronic orbitals, i.e. strongly and weakly
localized orbitals within the same atomic sphere rotate in different
directions under external strain.

\section{Acknowledgement}

This work was supported by the Nebraska Research Initiative, the
National Science Foundation and the Nanoelectronics Research
Initiative through the Materials Research Science and Engineering
Center at the University of Nebraska.

\end{document}